\documentclass[aps,prd,nofootinbib,twocolumn,superscriptaddress,floatfix]{revtex4}
\usepackage{mathrsfs}
\usepackage{latexsym}
\usepackage{amsmath}
\usepackage{amssymb}
\usepackage{graphicx}
\usepackage{longtable}
\usepackage{multirow}
\usepackage{bbm}
\usepackage{epsfig}
\usepackage[usenames]{color}
\usepackage[colorlinks=true,citecolor=blue,linkcolor=blue]{hyperref}
\allowdisplaybreaks

\begin{document}



\title{Hyperorder net baryon number fluctuations in nuclear matter at low temperature}

\author{Xin-ran Yang}
\affiliation{School of Physics, Xi'an Jiaotong University, Xi'an, 710049, China}
\affiliation{ MOE Key Laboratory for Nonequilibrium Synthesis and Modulation of Condensed Matter, Xi'an Jiaotong University, Xi'an, 710049, China}

\author{Guo-yun Shao}
\email[Corresponding author: ]{gyshao@mail.xjtu.edu.cn}
\affiliation{School of Physics, Xi'an Jiaotong University, Xi'an, 710049, China}
\affiliation{ MOE Key Laboratory for Nonequilibrium Synthesis and Modulation of Condensed Matter, Xi'an Jiaotong University, Xi'an, 710049, China}

\author{Wei-bo He} 
\affiliation{School of Physics, Peking University, Beijing, 100871, China}

\begin{abstract}
We calculate the density fluctuations of net baryon number up to sixth order induced by the interactions of nuclear matter, and explore their relationship with the nuclear liquid-gas phase transition (LGPT), including the stable and  metastable phase as well as the region far from the phase transition. The results show that dramatic density fluctuations exist in the vicinity of LGPT, and the hyperorder density fluctuations are more sensitive than the lower order ones to the interactions and structural properties of nuclear matter. The study also indicates that, even far away from the critical region of  LGPT, the hadronic interactions can still lead to significant hyperorder density fluctuations. In combination with the chemical freeze-out line fitted from the experimental data,  the derived results can be referred to investigate the chiral phase transition, nuclear LGPT, as well as the analysis of related experimental signals. 
\end{abstract}


\maketitle

\section{introduction}

Exploring the phase structure of strongly interacting matter is a crucial topic in high energy nuclear physics. One of the tasks is to investigate the types of phase transition from quark-gluon plasma to hadronic matter.  Studies based on the hadron resonance gas~~(HRG) model and lattice QCD show that the phase transition at high temperature and low chemical potential is a smooth crossover~\cite{Aoki06, Bazavov19, Borsanyi13,Bazavov14, Bazavov17,Borsanyi14,Borsanyi20}.  However, the application of lattice QCD in the high-density region is limited due to the sign problem associated with the fermion determinant. Nevertheless, most studies with  the effective  quark models~(e.g., \cite{Fukushima04,Ratti06,Costa10,Fu08, Sasaki12, Ferreira14, Shao2018, Schaefer10, Skokov11,Liu2018,Chen20,Zhao23}), the Dyson-Schwinger equation approach~\cite{Qin11,Gao16,Gao20, Fischer14,Shi14}, the functional renormalization group theory~\cite{Fu20,Rennecke17, Fu21},  suggest that the chiral phase transition in high chemical potential region undergoes a first-order phase transition.

It is proposed that the phase structure of QCD can be studied through the fluctuations and correlations of conserved charges~\cite{Stephanov}. The net proton~(proxy of net baryon) cumulants have been measured at the Relativistic Heavy Ion Collider (RHIC)~\cite{Aggarwal10, Adamczyk14}, and a nonmonotonic behavior in the fourth-order fluctuations as a function of collision energy was discovered~\cite{Luo2014, Luo2017}, which has sparked extensive discussions about the existence of the QCD critical endpoint~(CEP). Recent publication from  STAR collaboration further investigated the energy dependence of net proton number fluctuations up to sixth order, showing not only a nonmonotonic behavior but also more strongly hyperorder density fluctuations at lower collision energies~\cite{Aboona23}. Particularly, the fluctuations  of net proton number in collisions  at center-of mass energy $\sqrt{s_{NN}}=3\,$GeV is markedly different from those at $7.7\,$GeV and above. The study further suggests that in collisions at $\sqrt{s_{NN}}=3\,$GeV, the fluctuation distribution of net proton number is primarily dominated by the interaction among hadrons~\cite{Aboona23}.

The  new experimental data from RHIC have drawn further attention to the critical behavior of QCD phase transition and motivated the study of density fluctuations induced by the interaction among hadrons~\cite{Huang23, ShaoJ23,Marczenko23}. It is well-known that the interaction in nuclear matter can lead to the LGPT in the low-temperature region~\cite{Chomaz04,Pochodzalla95,Borderie01,Botvina95,Agostino99,Srivastava02,Elliott02}.  The critical point of the nuclear LGPT is estimated to be at a temperature of approximately 15 MeV and a chemical potential of around 923 MeV, although the values slightly depend on the  nuclear models. By far various properties of nuclear matter and the liquid-gas phase transition have been extensively studied in the literature~\cite{He23,Shao20,Chomaz04,Pochodzalla95,Borderie01,Botvina95,Agostino99,Srivastava02,Elliott02,Xu2023,Ma1999,Deng2022,Hempel13, Mukherjee17,Xu13,Savchuk20,Wang20}. 

The presence of strong density fluctuations in collisions at  $\sqrt{s_{NN}}=3\,$GeV raises the question of how the haddronic interaction and LGPT affect the hyperorder density fluctuations at lower-energy regimes. The authors in ~Ref.~\cite{Poberezhnyuk19,Shao20,Xu2023} calculated the density fluctuations up to  fourth order induced by hadronic interactions and analyzed the relationship between the lower-order density fluctuations and nuclear LGPT. In Ref.~\cite{Vovchenko17,Poberezhnyuk21}, a van der Waals model was taken to study the higher order fluctuation distributions of net baryon number in both the pure and mixed phase. In Ref.~\cite{Marczenko23}, the properties of the second-order susceptibility of net baryon number density for positive- and negative-parity nucleons were recently examined near the chiral and nuclear liquid-gas phase transitions using the double parity  model, in which both the chiral phase transition and nuclear LGPT are effectively included. However, a systematic study on the hyperorder density fluctuations in combination with the recent experimental data is still absent.

Considering the hyperorder density fluctuations dominated by the interactions among hadrons at lower energy regimes~(below $3\,$GeV) and the plan of the HADES collaboration at GSI Helmholtzzentrum für Schwerionenforschung to measure higher-order net proton and net charge fluctuations in central Au + Au reactions at collision energies ranging from $0.2A$ to $1.0A\,$GeV to probe the LGPT region~\cite{Bluhm20}, it is significant  to investigate the hyperorder density fluctuations in nuclear matter. In this study, we  will calculate the hyperorder net baryon distributions  up to sixth order using the nonlinear Walecka model.  The density fluctuations in the the stable and metastable phases of nucelar LGPT, as well as the area far from the critical region will be all explored. We will discuss the relations between the nucleon-nucleon interaction, the nucelar LGPT and the organization structure of the hyperorder density fluctuations of net baryon number. This investigation has a consequence for indicating the chiral phase transition, nuclear LGPT, as well as the analysis of related experimental signals.

The paper is organized as follows. In Sec.~II, we introduce the formulas to describe fluctuations of conserved charges and the nonlinear Walecka model in the mean-field approximation. In Sec.~III, we illustrate the numerical results of net-baryon number fluctuations, and discuss their relations with nucleon-nucleon interactions and the LGPT in nuclear matter. A summary is finally given in Sec. IV.

\section{ Fluctuations of conserved charges and the nonlinear Walecka model}

For a thermodynamic system, the fluctuations of conserved charges are sensitive observables  for  a phase transition, in particular, for a critical phenomenon.   The pressure of a system in the grand-canonical ensemble is related to the logarithm of the partition function \cite{Karsch15}:
\begin{equation}\label{}
\frac{P}{T^4}=\frac{1}{VT^3}\ln[Z(V,T,\mu_B, \mu_Q, \mu_S)],
\end{equation}  
where $\mu_B, \mu_Q, \mu_S$ are the chemical potentials of conserved charges, the baryon number, electric charge and strangeness in strong interaction, respectively. The generalized susceptibilities can be derived by taking the partial derivatives of the pressure with respect to the corresponding chemical potentials \cite{Luo2017}
\begin{equation}\label{}
\chi^{BQS}_{ijk}=\frac{\partial^{i+j+k}[P/T^4]}{\partial(\mu_B/T)^i \partial(\mu_Q/T)^j \partial(\mu_S/T)^k}.
\end{equation}  

The cumulants of multiplicity distributions of the conserved charges are connected with the generalized susceptibilities by
 \begin{equation}\label{Crho}
\!C^{BQS}_{ijk}\!=\!\frac{\partial^{i\!+\!j\!+\,k}\ln[Z(V,T,\mu_B, \mu_Q, \mu_S)]}{\partial(\mu_B/T)^i \partial(\mu_Q/T)^j \partial(\mu_S/T)^k}\!=\!V\!T^3\chi^{BQS}_{ijk}\, .
\end{equation} 
To eliminate the volume dependence in heavy-ion collision experiments, observables are usually constructed by the ratio of cumulants, and then can be compared with theoretical calculations of the generalized susceptibilities with
 \begin{equation}\label{}
\frac{C^{BQS}_{ijk}}{C^{BQS}_{lmn}} =\frac{\chi^{BQS}_{ijk}}{\chi^{BQS}_{lmn}}.
 \end{equation} 

The nonlinear Walecka model is taken to calculate fluctuations of net baryon number in nuclear matter. This model is generally used to describe  the properties of finite nuclei and the equation of state of nuclear matter.  The study in Ref.~\cite{Fukushima15} also indicates the approximate equivalence of this model to the hadron resonance gas model at low temperature and small density.
The Lagrangian density for nucleons-meson system in the nonlinear Walecka model \cite{Glendenning97} is 
\begin{eqnarray}\label{lagrangian}
\cal{L}\!&\!=\!&\sum_N\bar{\psi}_N\!\big[i\gamma_{\mu}\partial^{\mu}\!-\!(\!m_N
          \!-\! g_{\sigma }\sigma)\!
                 \! -\!g_{\omega }\gamma_{\mu}\omega^{\mu} \!-\! g_{\rho }\!\gamma_{\mu}\boldsymbol\tau_{}\!\cdot\!\boldsymbol
\rho^{\mu} \big]\!\psi_N       \nonumber\\
         & &    +\frac{1}{2}\left(\partial_{\mu}\sigma\partial^
{\mu}\sigma-m_{\sigma}^{2}\sigma^{2}\right)\!-\! \frac{1}{3} bm_N\,(g_{\sigma} \sigma)^3-\frac{1}{4} c\,
(g_{\sigma} \sigma)^4
                    \nonumber\\
       & &+\frac{1}{2}m^{2}_{\omega} \omega_{\mu}\omega^{\mu}
          -\frac{1}{4}\omega_{\mu\nu}\omega^{\mu\nu}  \nonumber \\
& &    +\frac{1}{2}m^{2}_{\rho}\boldsymbol\rho_{\mu} \! \cdot \! \boldsymbol
\rho^{\mu}   \!- \! \frac{1}{4}\boldsymbol\rho_{\mu\nu} \! \cdot \! \boldsymbol\rho^{\mu\nu} ,
 \end{eqnarray}
 where $
\omega_{\mu\nu}= \partial_\mu \omega_\nu - \partial_\nu
\omega_\mu$, $ \rho_{\mu\nu} =\partial_\mu
\boldsymbol\rho_\nu -\partial_\nu \boldsymbol\rho_\mu$. The interactions between
baryons are mediated by $\sigma,\,\omega,\,\rho$ mesons.

 The thermodynamical  potential of the nucleons-meson system can be derived in the mean-field approximation as
\begin{eqnarray}
 \Omega\!&\!=&\!-\!\beta^{-1} \sum_{N} 2\! \int \frac{d^{3} \boldsymbol{k}}{(2 \pi)^{3}}\!\bigg[\ln \!\left(1+e^{-\beta\left(E_{N}^{*}(k)-\mu_{N}^{*}\!\right)}\right)\! \nonumber \\ 
 &&+\!\ln\! \left(1\!+\!e^{-\beta\left(E_{N}^{*}(k)+\mu_{N}^{*}\right)}\right)\!\bigg]\! +\!\frac{1}{2} m_{\sigma}^{2} \sigma^{2}\!+\!\frac{1}{3} b m_{N}\left(g_{\sigma} \sigma\right)^{3} \nonumber \\
 &&+\frac{1}{4} c\left(g_{\sigma} \sigma\right)^{4}-\frac{1}{2} m_{\omega}^{2} \omega_{}^{2}-\frac{1}{2} m_{\rho}^{2} \rho_{3}^{2}  ,
  \end{eqnarray}
where $\beta=1/T$, $E_{N}^{*}=\sqrt{k^{2}+m_{N}^{*2}}$,  and  $\rho_{3}$ is the third component of $\rho$ meson field.  The effective chemical potential $\mu_{N}^{*}$ is defined as  $\mu_{N}^{*}=\mu_{N}-g_{\omega} \omega_{}-\tau_{3 N} g_{\rho} \rho_{3}$ ($\tau_{3 N}=1/2$ for proton, $-1/2$ for neutron).

By minimizing the thermodynamical potential
\begin{equation}
\frac{\partial \Omega}{\partial \sigma}=\frac{\partial \Omega}{\partial \omega_{}}=\frac{\partial \Omega}{\partial \rho_{3}}=0,
\end{equation}
the meson field equations can be derived as
\begin{equation}\label{sigma}
g_{\sigma} \sigma\!=\!\left(\frac{g_{\sigma}}{m_{\sigma}}\right)^{2}\!\left[\rho_{p}^{s}+\rho_{n}^{s}-b m_{N}\left(g_{\sigma} \sigma\right)^{2}-c\left(g_{\sigma} \sigma\right)^{3}\right],
\end{equation}
\begin{equation}
g_{\omega} \omega=\left(\frac{g_{\omega}}{m_{\omega}}\right)^{2}\left(\rho_{p}+\rho_{n}\right),
\end{equation}
\begin{equation}\label{rho}
g_{\rho} \rho_{3}=\frac{1}{2}\left(\frac{g_{\rho}}{m_{\rho}}\right)^{2} \left(\rho_{p}-\rho_{n}\right).
\end{equation}

In Eqs.(\ref{sigma})-(\ref{rho}), the nucleon number density
\begin{equation}
\rho_i=2 \int \frac{d^{3} \boldsymbol{k}}{(2 \pi)^{3}} [f\left(E_{i}^{*}-\mu_{i}^{*}\right)-\bar{f}\left(E_{i}^{*}+\mu_{i}^{*}\right) ],
\end{equation}
and the scalar density
\begin{equation}
\rho_{i}^{s}=2\int \frac{d^{3}\boldsymbol{k}}{(2 \pi)^{3}} \frac{m_{i}^{*}}{E_{i}^{*}}[f\left(E_{i}^{*}-\mu_{i}^{*}\right)+\bar{f}\left(E_{i}^{*}+\mu_{i}^{*}\right)],
\end{equation}
where $f(E_{i}^{*}-\mu_{i}^{*})$ and $\bar{ f} (E_{i}^{*}+\mu_{i}^{*})$ are the  fermion and antifermion distribution functions.

For a given temperature and chemical potential (or baryon number density), the meson field equations can be solved. 
In the present study,  the symmetric nuclear matter is considered to give an essential description of the density fluctuations. The model parameters, $g_\sigma, g_\omega, g_\rho, b$ and $ c$, are fitted with the compression modulus $K=240\,$MeV, the symmetric energy  $a_{sym}=31.3\,$MeV, 
the effective nucleon mass $m^*_N=m_N- g_\sigma \sigma=0.75m_N$~($m_N$ is the nucleon mass in vacuum) and the
 binding energy $B/A=-16.0\,$MeV at nuclear saturation density with $\rho_0=0.16\, fm^{-3}$. 

\section{numerical results and discussions}

\begin{figure}[htbp]
	\begin{center}
		\includegraphics[scale=0.42]{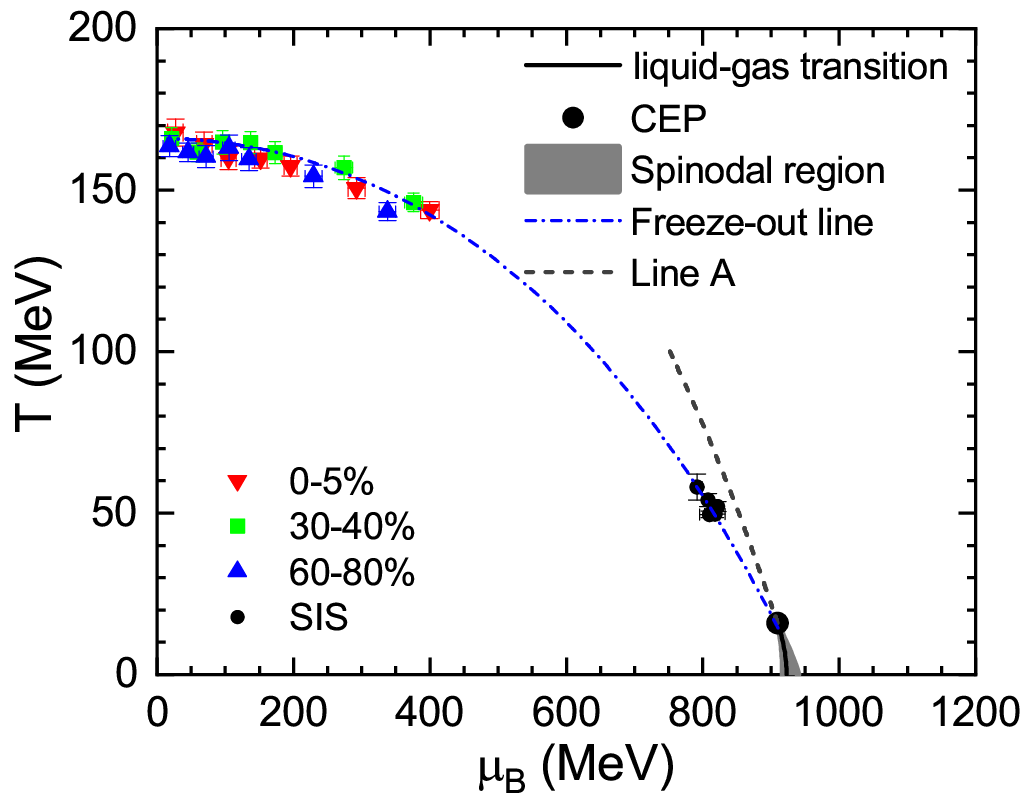}
		\caption{\label{fig:1}  Phase diagram of LGPT in nuclear matter, extracted chemical freeze-out temperature versus baryon chemical potential from STAR experiments~\cite{Adamczyk17} and SIS data~\cite{Cleymans99,Becattini01,Averbeck03}, as well as the the fitted chemical freeze-out curve from Ref.~\cite{Cleymans06}. ``Line A " is derived with  $\partial \sigma / \partial \mu_B$ taking the maximum value for a given temperature.}
	\end{center}
\end{figure}

We first plot in Fig.~\ref{fig:1} the phase diagram of nuclear matter in the temperature-chemical potential plane. The black solid line is the first-order liquid-gas phase transition line with a critical endpoint at $T_c=15.9\,$MeV and $\mu_c=909.8\,$MeV. 
The black dashed curve, marked with ``Line A ", is derived with  $\partial \sigma / \partial \mu_B$ taking the maximum value for a given temperature, 
which is connected to the nuclear liquid-gas phase transition. This plot is somewhat similar to the chiral crossover transformation line and first-order phase transition in quark model. They both correspond to the fastest change or jump  of thermodynamic order parameter~($\sigma$ field in Walecka model and chiral condensate in quark model) related to fermion mass. And the subsequent numerical results indeed indicate a similar structure of density fluctuations to QCD phase transition. The fluctuation distributions near the ``line A" above the critical temperature are most dramatic. These characteristics reveal essentially similar physical properties of the two phase transitions because they belong to the same universal class.

\begin{figure}[htbp]
	\begin{center}
		\includegraphics[scale=0.42]{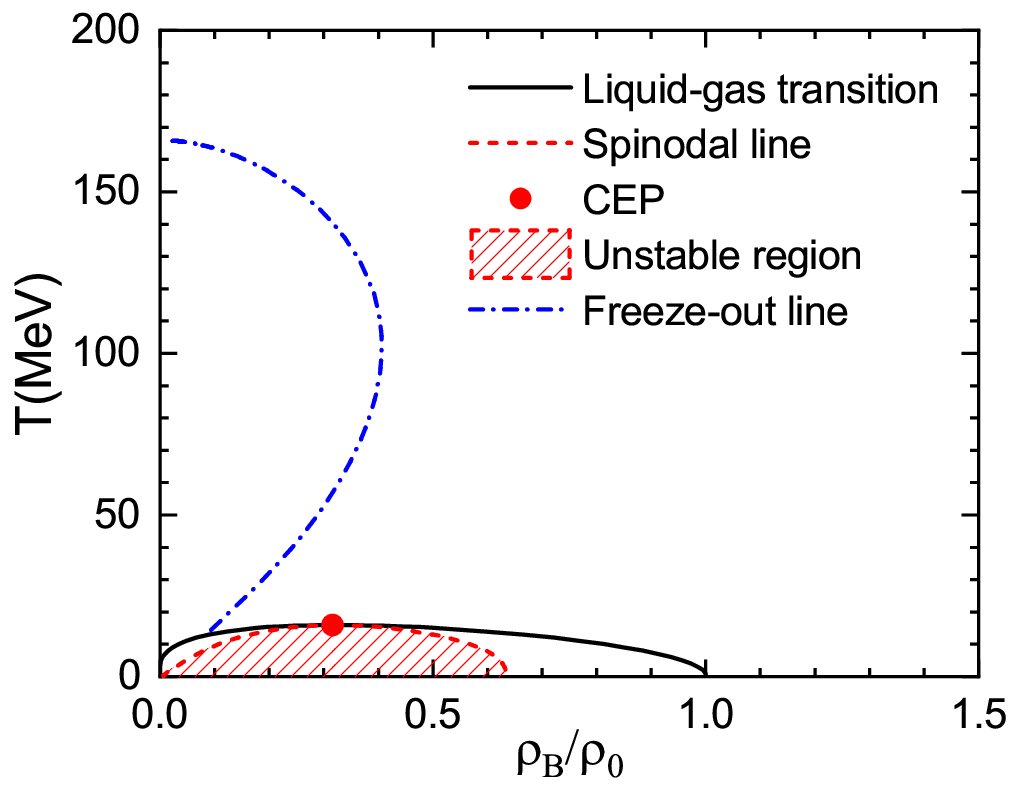}
		\caption{\label{fig:2} Phase diagram of LGPT  and the chemical freeze-out line in the $T-\rho_B$ plane derived in the nonlinear Walecka model on the basis of the chemical freeze-out line shown in Fig.~\ref{fig:1} with the fitted formula in Eq.(\ref{fl}).
		}
	\end{center}
\end{figure}

In Fig.~\ref{fig:1}, we also plot the fitted chemical freeze-out curve and the data from heavy-ion collision experiments~\cite{Cleymans06} for the convenience of later discussion of experimental signals. The fitted chemical freeze-out line can be described with 
\begin{equation}\label{fl}
	T\left(\mu_B\right)=a-b \mu_B^2-c \mu_B^4,
\end{equation}
where $a=0.166\, \mathrm{GeV}, b=0.139 \, \mathrm{GeV}^{-1}$ and $c=0.053 \, \mathrm{GeV}^{-3}$.

We plot in Fig.~\ref{fig:2} the phase diagram of liquid-gas transition in the temperature-density plane. 
The spinodal line is given with a red dashed curve. The area under the spinodal line is the unstable phase. The region between the spinodal line and the first-order phase transition line is the metastable phase. 
We also plot the chemical freeze-out curve, which will be referenced to discuss the density fluctuations in the  $T-\rho_B$ plane. The freeze-out line  in Fig.~\ref{fig:2} is translated from the curve in Fig.~\ref{fig:1} based on the calculation in the nonlinear Walecka model. Using the fitted relation of $T$ and $\mu_B$ at freeze-out given by Eq.(\ref{fl}), we can calculate the baryon density by solving Eqs.(\ref{sigma})-(\ref{rho}). A distinct feature is that the baryon density on the chemical freeze-out line is nonmonotonic with the decrease of collision energy~(or temperature). This behavior has been confirmed by the experimental data in combination with hadron resonance gas model~\cite{Randrup06, Poberezhnyuk19}. A similar result was also found in the $\sigma-\omega$ model in Ref.~\cite{Fukushima15}.

Besides, Fig.~\ref{fig:2} shows that the freeze-out baryon number density is close to zero near $T=0$. This is consistent with the experimental data mentioned above as well as the theoretic prediction, because the chemical freeze-out at low temperature should occur in the gas phase of the first-order LGPT of nuclear matter. From Fig.~\ref{fig:2}, it can be seen that the baryon density in the gas phase is much smaller than that in the liquid phase.  As a matter of fact, the baryon number densities at chemical freeze-out derived from experiments with  $\sqrt{s_{NN}}=200-3\,$GeV are all smaller than  the nuclear saturation density $\rho_0$~\cite{Randrup06}.

\begin{figure}[htbp]
  	\begin{center}
  		\includegraphics[scale=0.4]{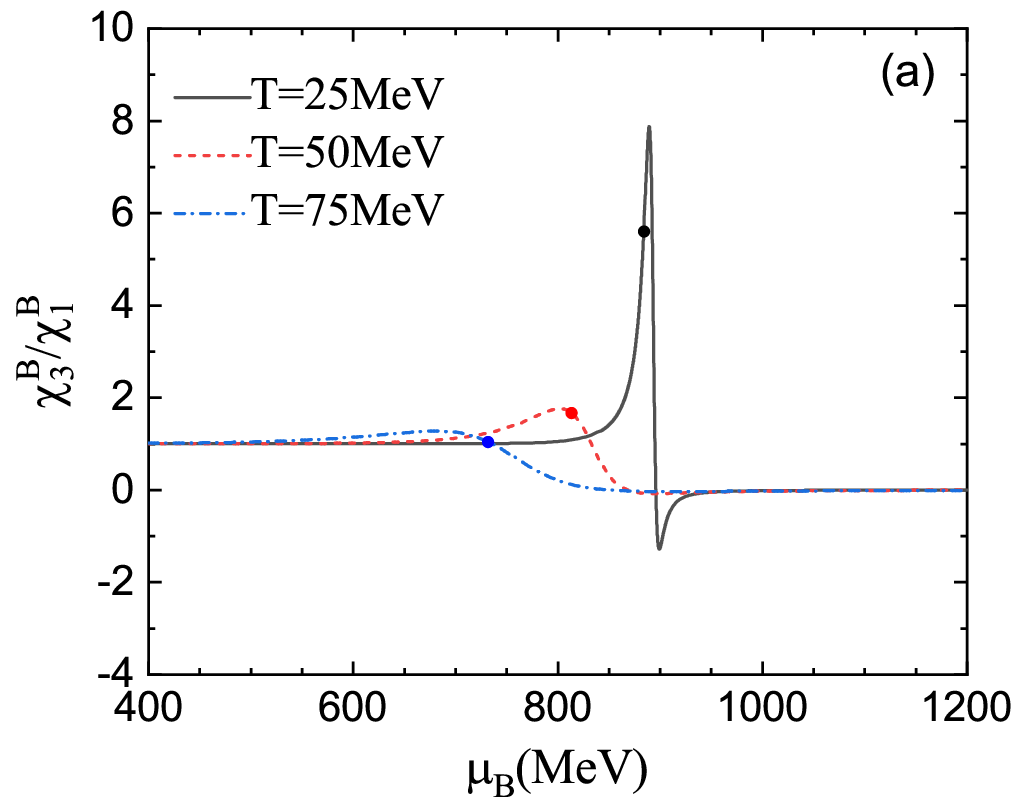}
  		\includegraphics[scale=0.4]{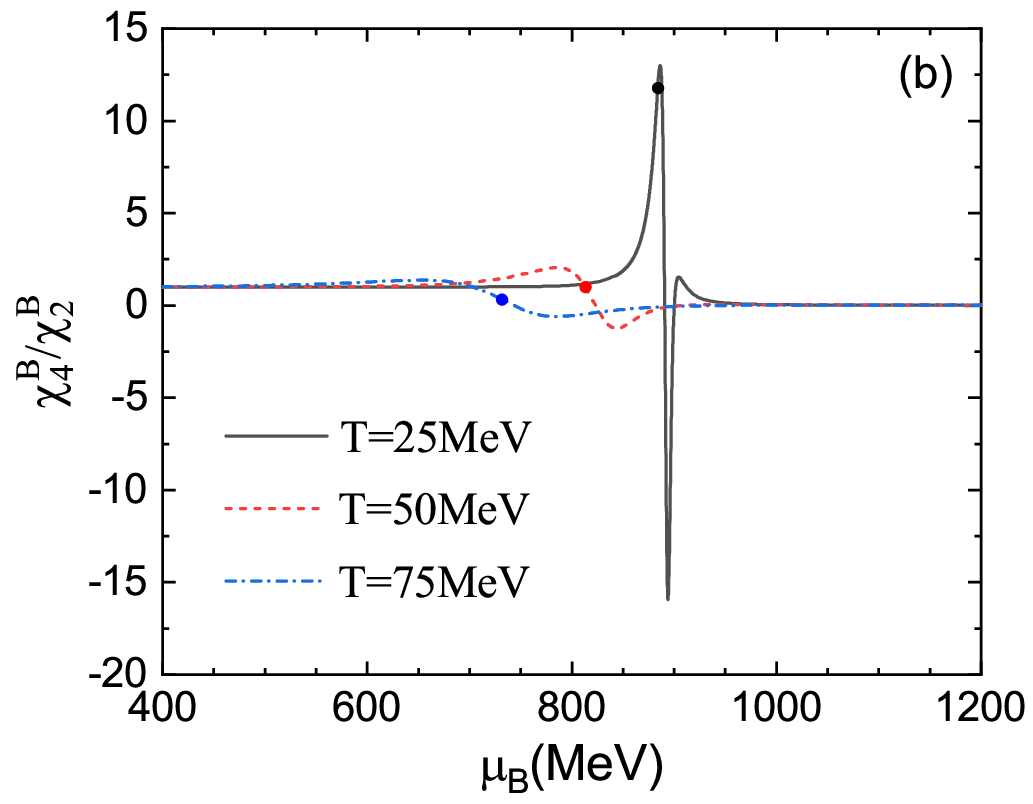}
  		\includegraphics[scale=0.4]{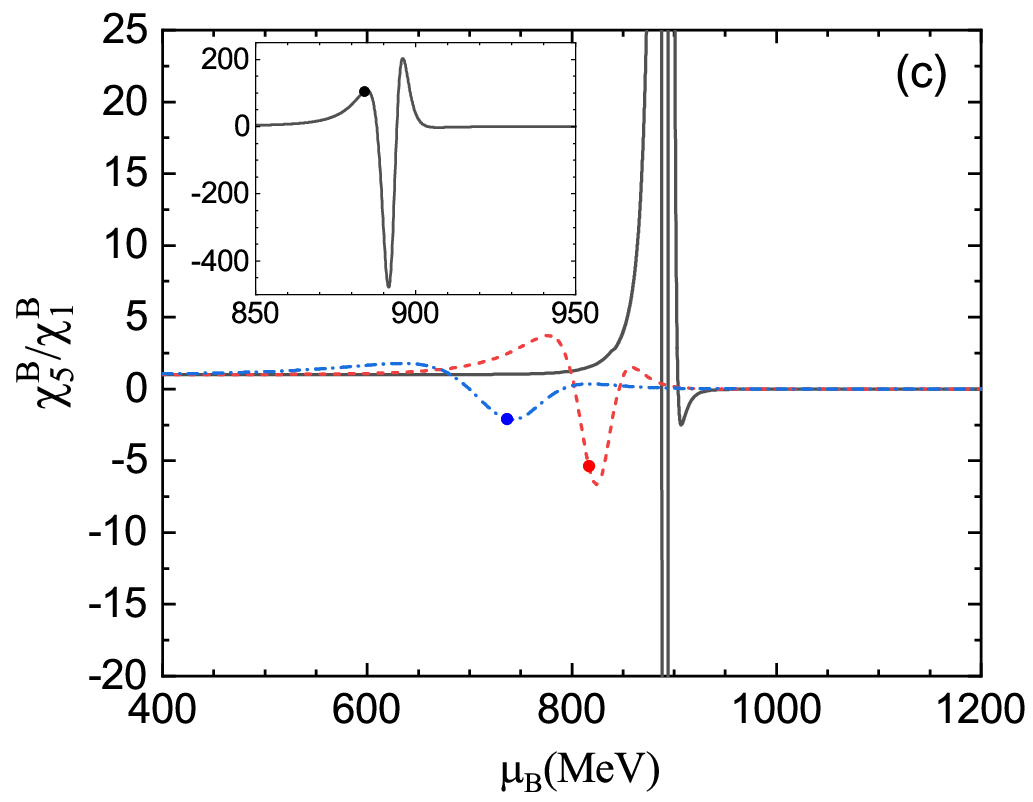}
  		\includegraphics[scale=0.4]{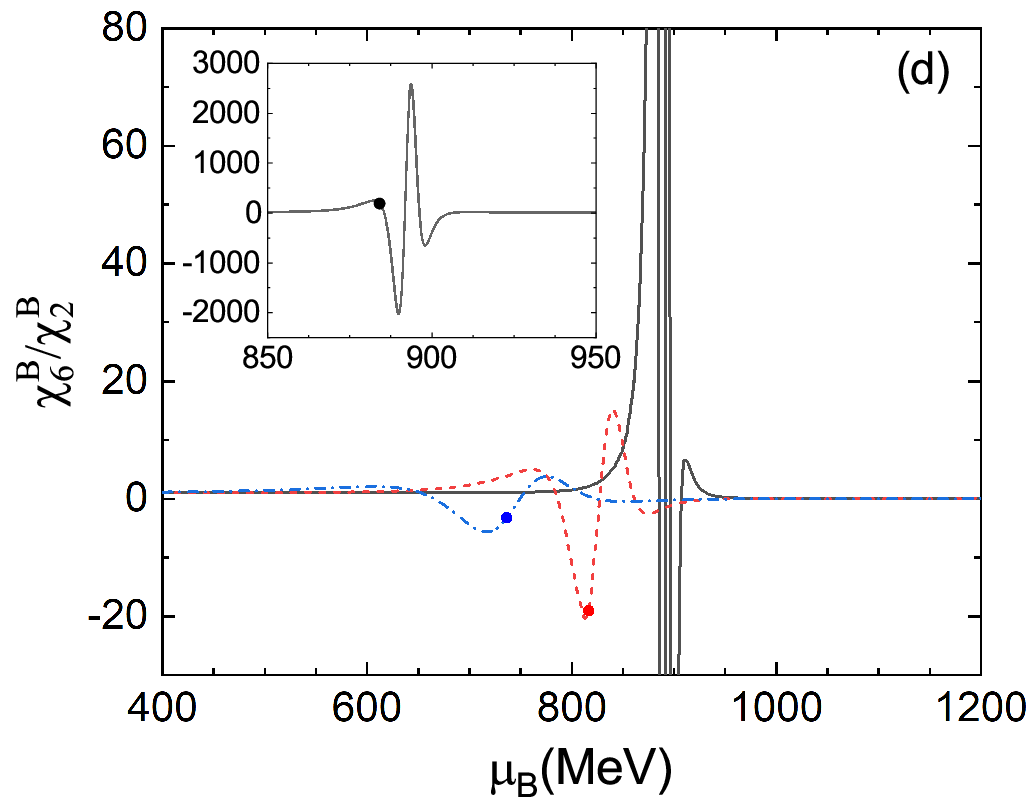}
  		\caption{\label{fig:chimu} Ratios of net baryon number susceptibilities as functions of chemical potential for different temperatures. Panel (a): $\chi_3^B/\chi_1^B$; Panel (b): $\chi_4^B/\chi_2^B$(kurtosis); Panel (c): $\chi_5^B/\chi_1^B$(hyperskewness); Panel (d): $\chi_6^B/\chi_2^B$(hyperkurtosis). The solid dots demonstrate the density fluctuations at chemical freeze-out given in Fig.~\ref{fig:1}.
  		}
  	\end{center}
  \end{figure}

We demonstrate in Fig.~\ref{fig:chimu} the ratios of susceptibilities, $\chi_3^B/\chi_1^B$, $\chi_4^B/\chi_2^B$, $\chi_5^B/\chi_1^B$ and $\chi_6^B/\chi_2^B$, as functions of baryon chemical potential for $T=75, 50, 25\,$MeV. 
The numerical results show that all these ratios of susceptibilities  change from one to zero as the chemical potential increases. Such a behavior is analogous to the density fluctuations of chiral phase transition from a chiral broken phase to a chiral restored phase~\cite{Shao2018}. The dynamical nucleon mass plays a similar role for nuclear matter.
 The strong density fluctuation for each temperature are roughly located in the region where $\partial \sigma / \partial \mu_B$ changes rapidly.
 The collection of maxima of $\partial \sigma / \partial \mu_B$ at different temperatures above the critical endpoint corresponds to the  dashed line A plotted in Fig.~\ref{fig:1}.
  
  \begin{figure}[htbp]
  	\begin{center}
  		\includegraphics[scale=0.4]{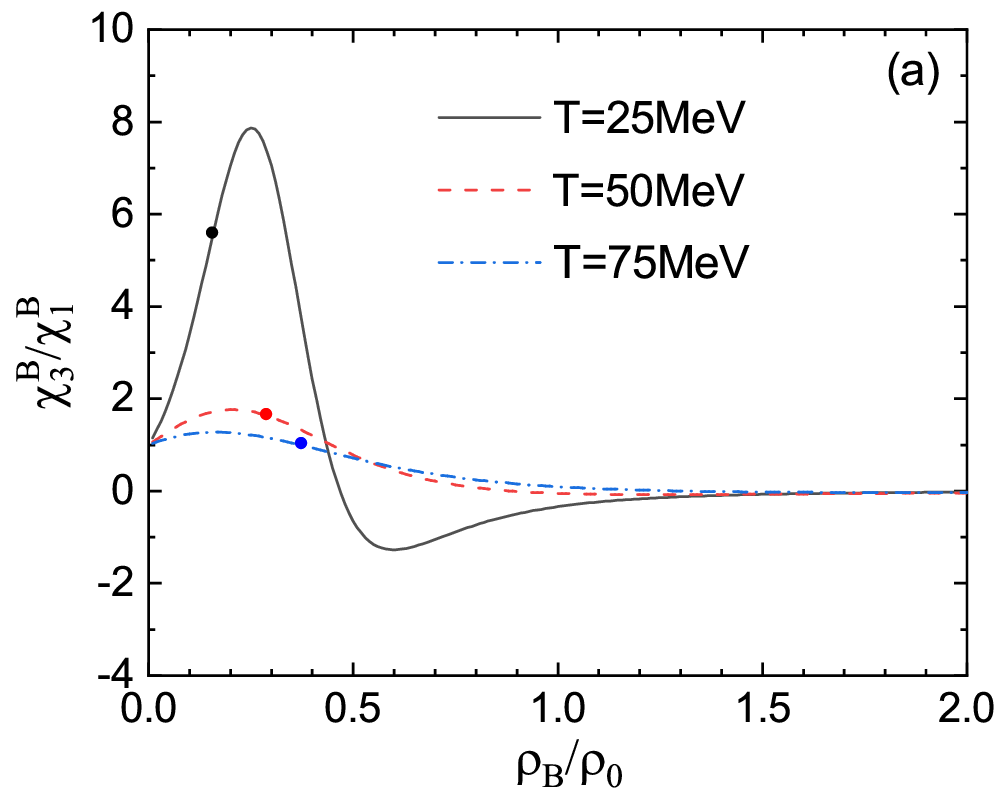}
  		\includegraphics[scale=0.4]{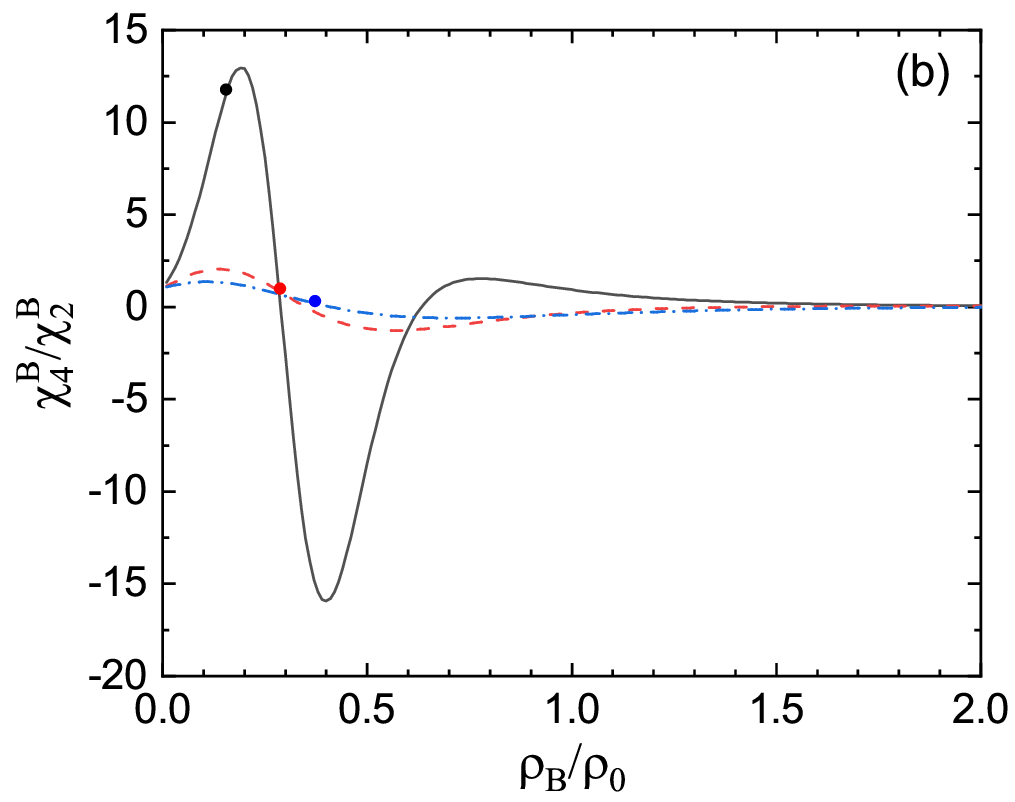}
  		\includegraphics[scale=0.4]{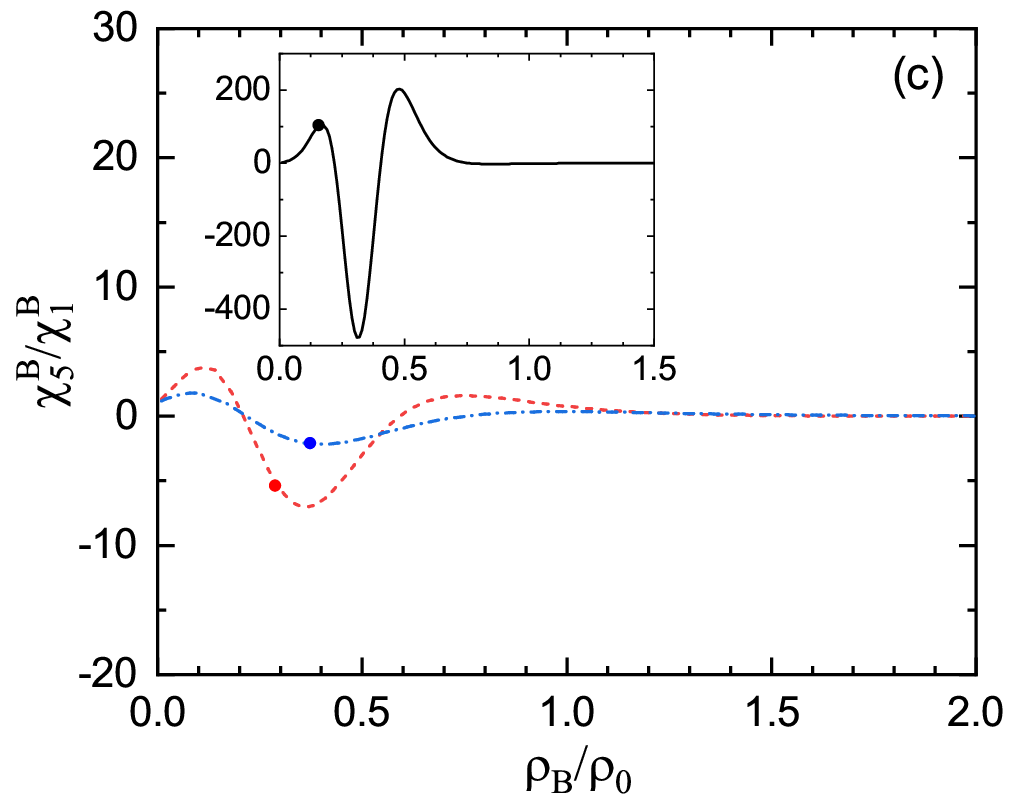}
  		\includegraphics[scale=0.4]{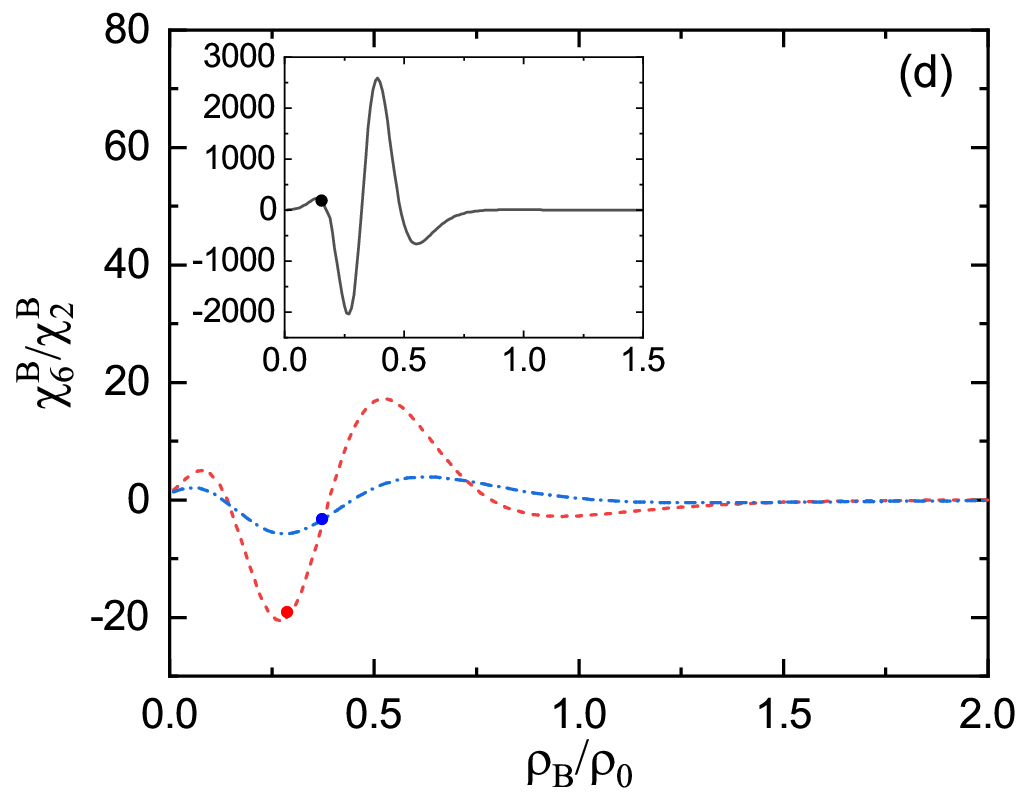}
  		\caption{\label{fig:chirho}Ratios of net baryon number susceptibilities  as functions of  baryon density for different temperatures. Panel (a): $\chi_3^B/\chi_1^B$; Panel (b): $\chi_4^B/\chi_2^B$(kurtosis); Panel (c): $\chi_5^B/\chi_1^B$(hyperskewness); Panel (d): $\chi_6^B/\chi_2^B$(hyperkurtosis). The solid dots demonstrate the density fluctuations at chemical freeze-out given in Fig.~\ref{fig:1}.
  		}
  	\end{center}
  \end{figure}

Fig.~\ref{fig:chimu} also shows that  the amplitude of density fluctuations is relatively small at $T=75\,$MeV. As the temperature decreases and approaches the critical region, the density fluctuations gradually increase and diverge at the CEP. This  indicates a close relationship between the density fluctuations of conserved charges and the phase transition of nuclear matter. From Fig.~\ref{fig:chimu}, we can  see that the hyperorder density fluctuations~(hyperskewness, $\chi_5^B/\chi_1^B$ and hyperkurtosis, $\chi_6^B/\chi_2^B$) are dramatic than those of lower orders~($\chi_3^B/\chi_1^B$ and $\chi_4^B/\chi_2^B$), meaning that the higher-order fluctuations are more sensitive to observe in experiments. Similar behaviors aroused by the chiral phase transition have been found in the beam energy scan experiments  at RHIC STAR~\cite{Aboona23}. Additionally, we can observe  from Fig.~\ref{fig:chimu} that $\chi_3^B/\chi_1^B$ and $\chi_4^B/\chi_2^B$ have one maximum and one minimum, while  $\chi_5^B/\chi_1^B$~($\chi_6^B/\chi_2^B$) has two maximums and one minimum~(two minimums).

The solid dots on the curves in Fig.~\ref{fig:chimu} demonstrate the fluctuation distributions at the points of chemical freeze-out, corresponding to the chemical freeze-out line plotted in Fig.~\ref{fig:1}. It can be observed that $\chi_3^B/\chi_1^B$ and $\chi_4^B/\chi_2^B$ are all positive for $T=75, 50$ and $25\,$MeV at chemical freeze-out. However,  $\chi_5^B/\chi_1^B$ and $\chi_6^B/\chi_2^B$ are negative at $T=75$ and $50\,$MeV, and  become positive at $T=25\,$MeV. Note that the negative $\chi_5^B/\chi_1^B$ and $\chi_6^B/\chi_2^B$ also appear at chemical freeze-out at higher temperatures for  QCD phase transition~\cite{Aboona23}. Interestingly, it is found that the ratios of net proton number distributions, $C_4/C_2$, $C_5/C_1$ and $C_6/C_2$ at $\sqrt{s_{NN}}=3\,$GeV with the collision centrality $0\%-40\%$, are very close to the  numerical results of  $\chi_4^B/\chi_2^B$, $\chi_5^B/\chi_1^B$ and $\chi_6^B/\chi_2^B$ at $T=28\,$MeV and $\mu_B=864\,$MeV in the nonlinear Walecka model.  However, the physical conditions, $T=28\,$MeV and $\mu_B=864\,$MeV, are different from the chemical freeze-out condition derived in  collisions at a center-of-mass energy of $3\,$GeV. For the later case,  $T \approx 70\,$MeV and $\mu_B\approx750\,$MeV are extracted at the chemical freeze-out. 
We also notice that there is a different argument  based on the data of only the net proton cumulants $C_4/C_2$ with  collision centralities $0\%-5\%$~\cite{Xu2023}. Higher accuracy data of  hyperskewness and hyperkurtosis at lower collision energies are needed to distinguish between the two scenarios.

We further plot the ratios of net baryon number susceptibilities in the temperature-density plane in Fig.~\ref{fig:chirho}. Compared with the results in Fig.~\ref{fig:chimu}, the distributions of density fluctuations appear to be broadened in the $T-\rho_B$ panel. The solid dots in Fig.~\ref{fig:chirho} indicate that the baryon density at chemical freeze-out decreases as the temperature descends from $75$ to $25\,$MeV. In contrast, Fig.~\ref{fig:chimu} shows that the chemical potential at  freeze-out increases as the temperature decreases from $75$ to $25\,$MeV. These results are consistent with the chemical freeze-out curve given in Figs.~\ref{fig:1} and 2, since the variation of baryon density along the chemical freeze-out line is nonmonotonic in the temperature-density phase diagram.

 \begin{figure*} [htbp]
	\centering
	\includegraphics[width=.45\linewidth]{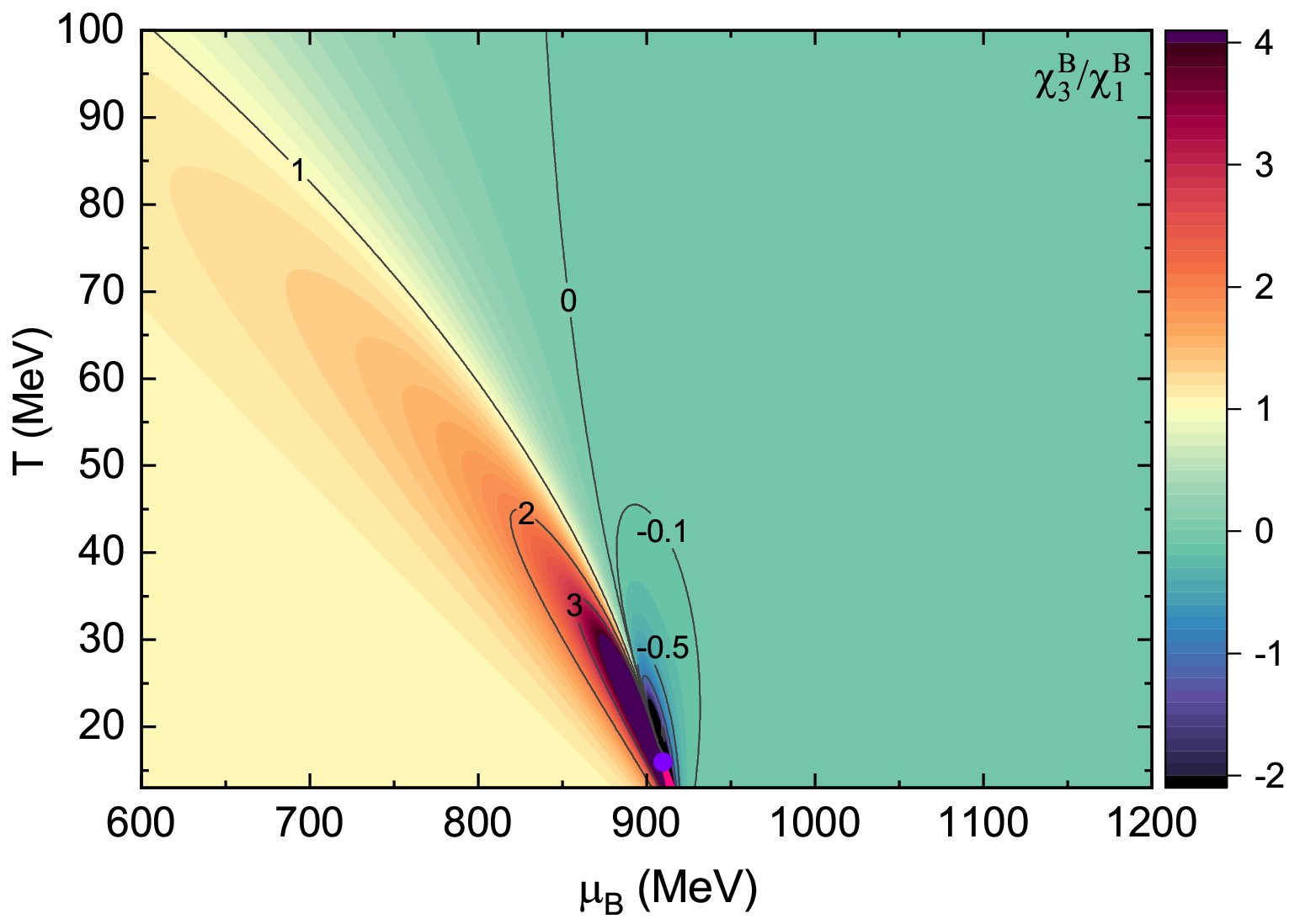}\includegraphics[width=.45\linewidth]{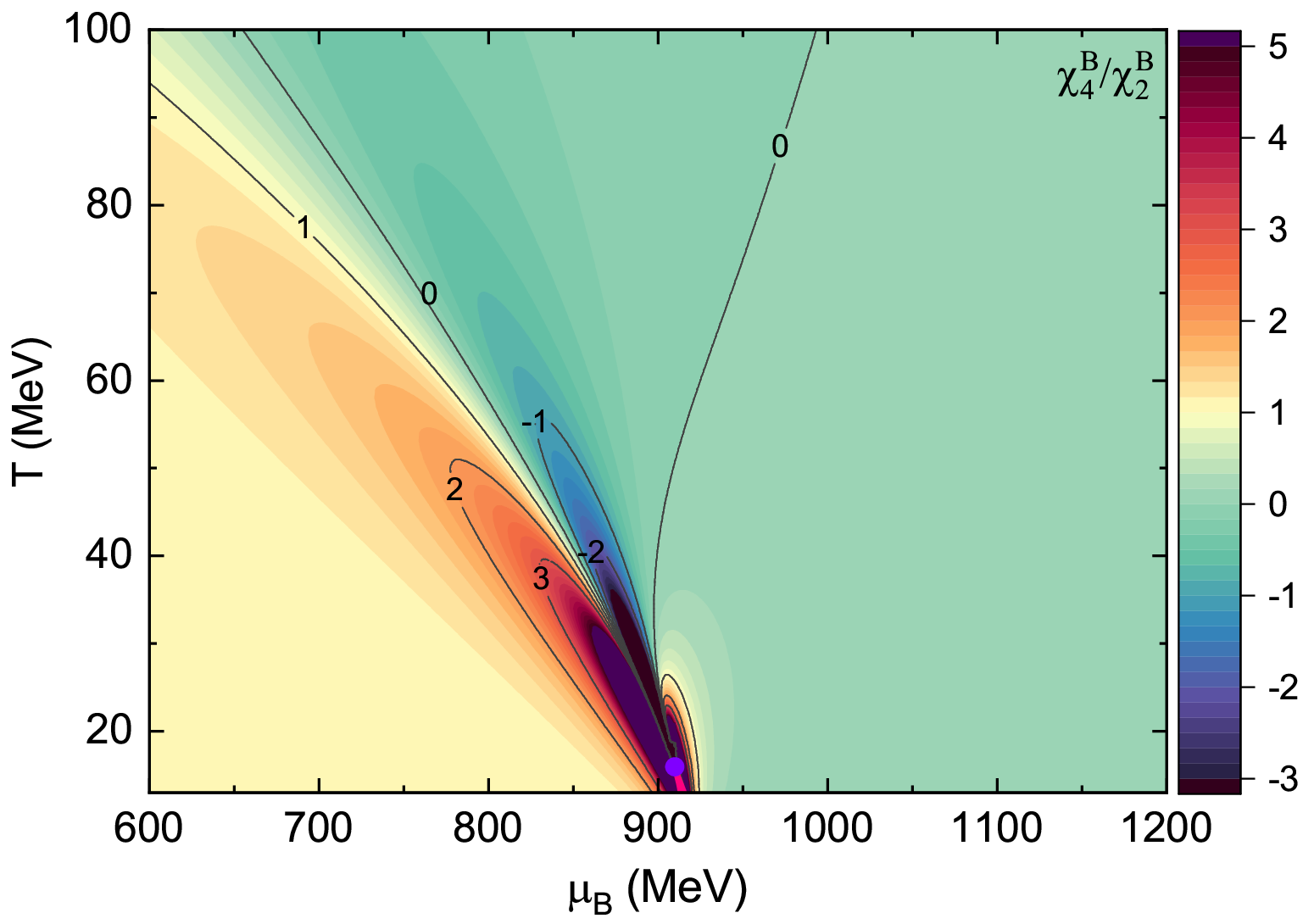}
	\includegraphics[width=.45\linewidth]{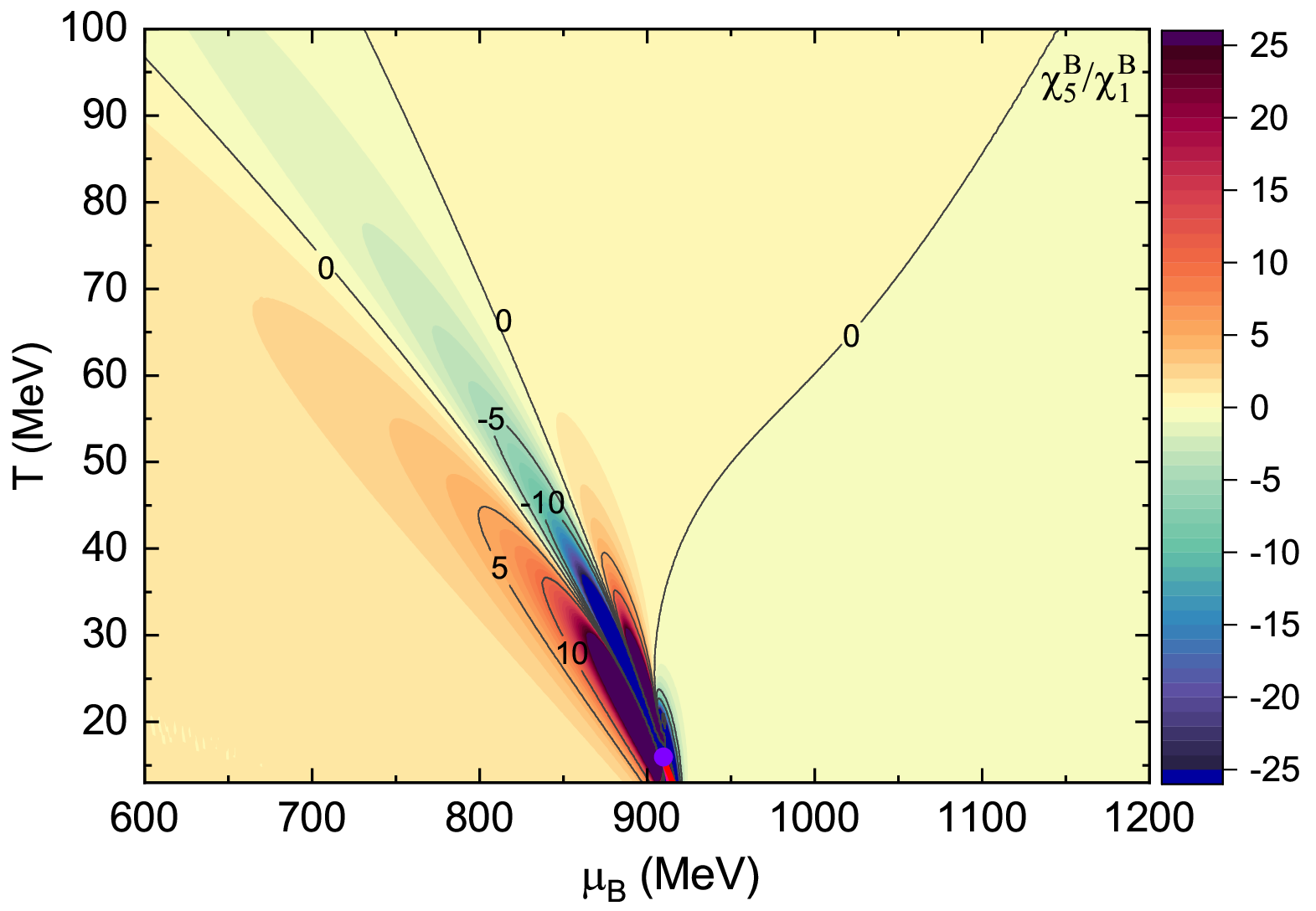}\includegraphics[width=.45\linewidth]{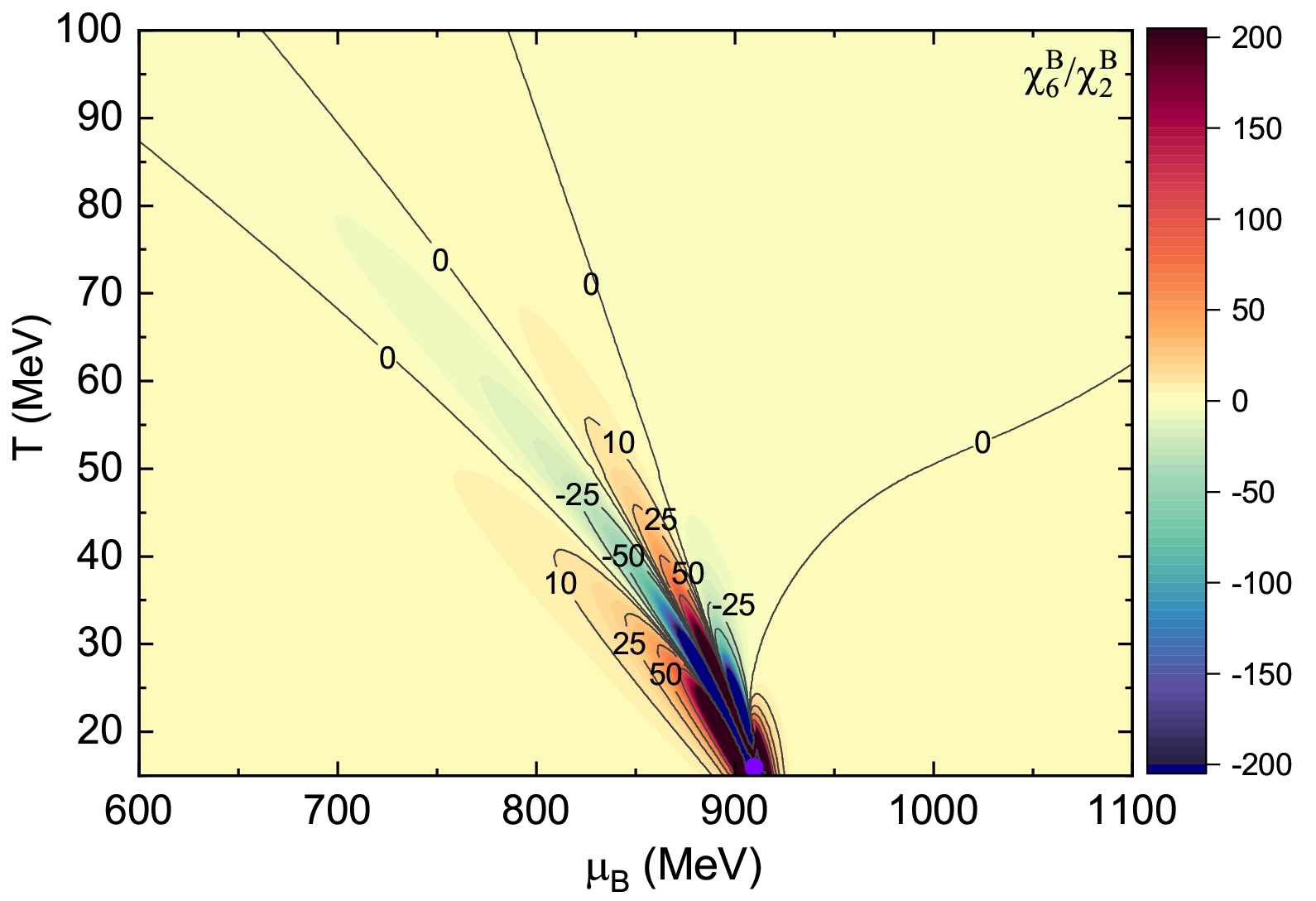}
	\caption{(color online) Contour plots of the ratios of net baryon number susceptibilities, $\chi_3^B/\chi_1^B$, $\chi_4^B/\chi_2^B$, 
		$\chi_5^B/\chi_1^B$ and $\chi_6^B/\chi_2^B$, in the $T-\mu_B$ phase diagram.
	} 
	\label{fig:5}
\end{figure*} 

\begin{figure*} [htbp]
\centering
\includegraphics[width=.45\linewidth]{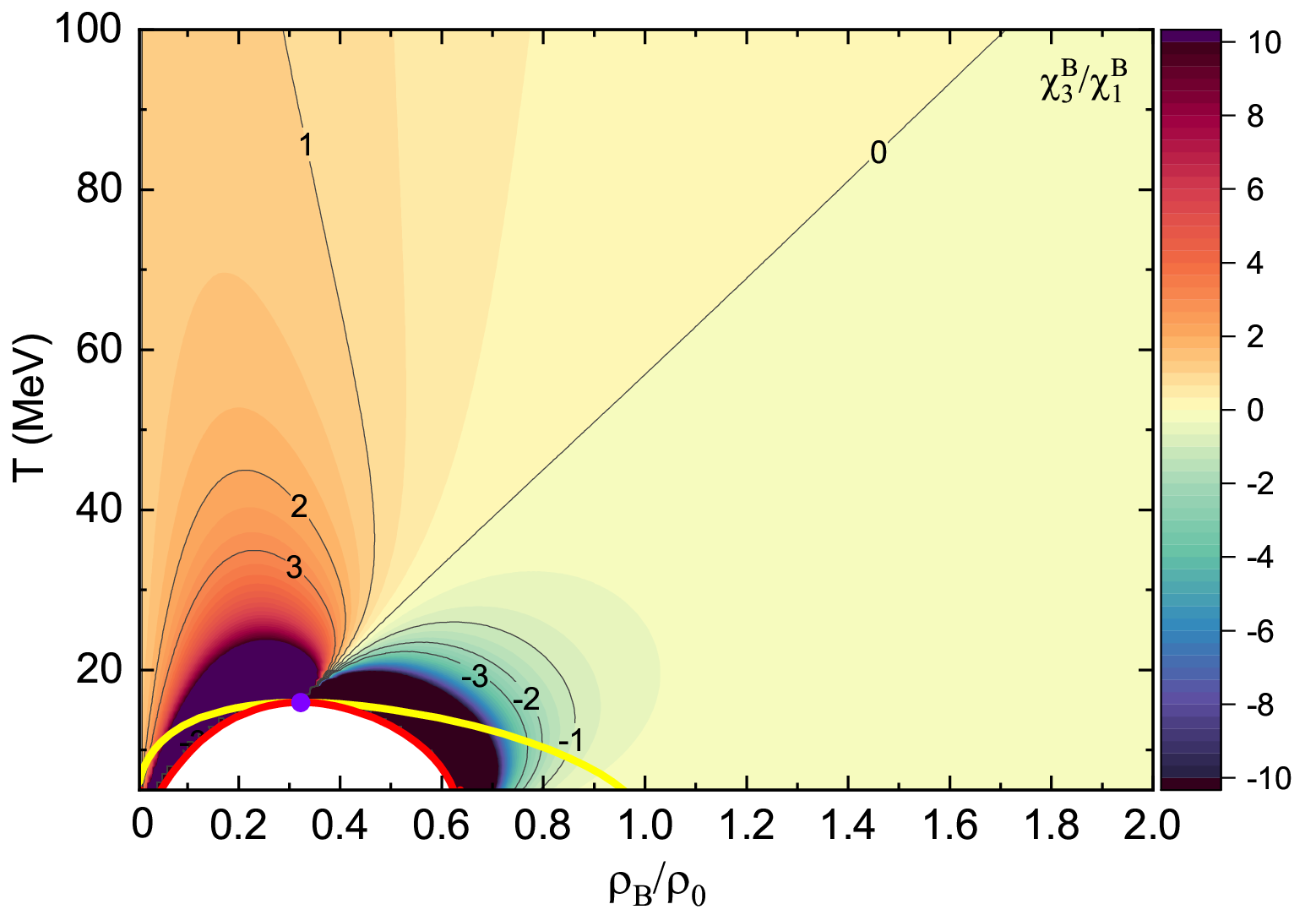}\includegraphics[width=.45\linewidth]{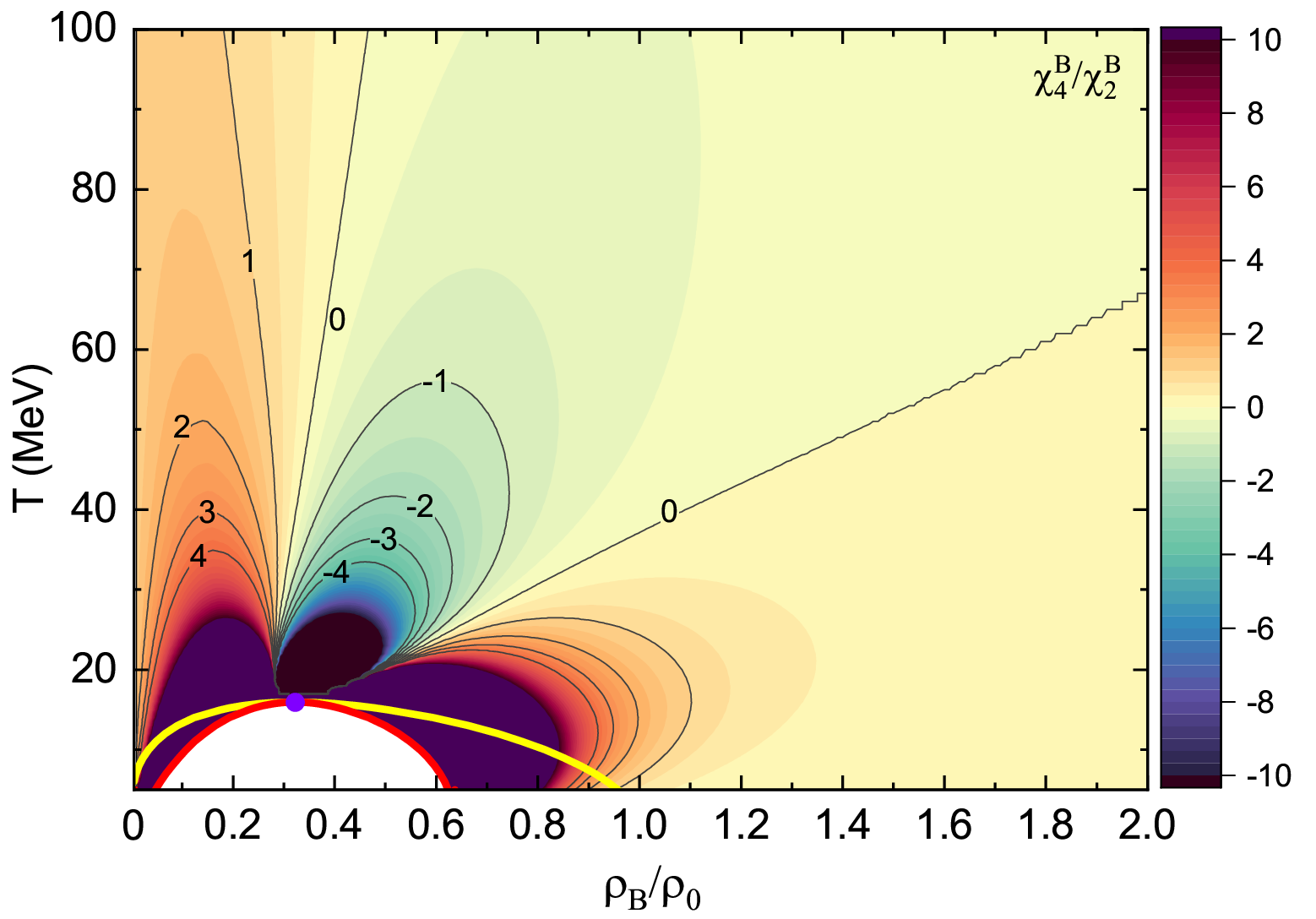}
\includegraphics[width=.45\linewidth]{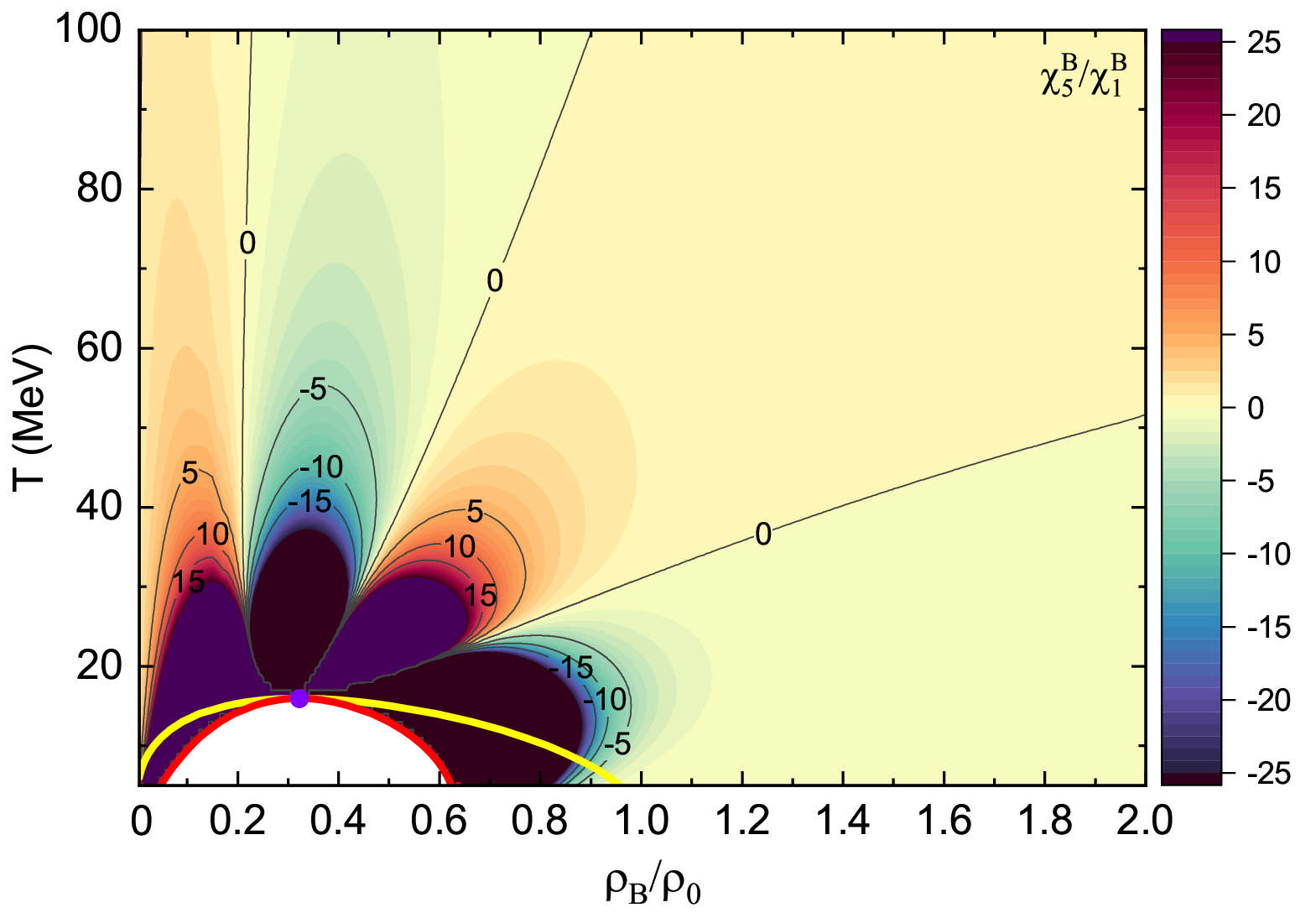}\includegraphics[width=.45\linewidth]{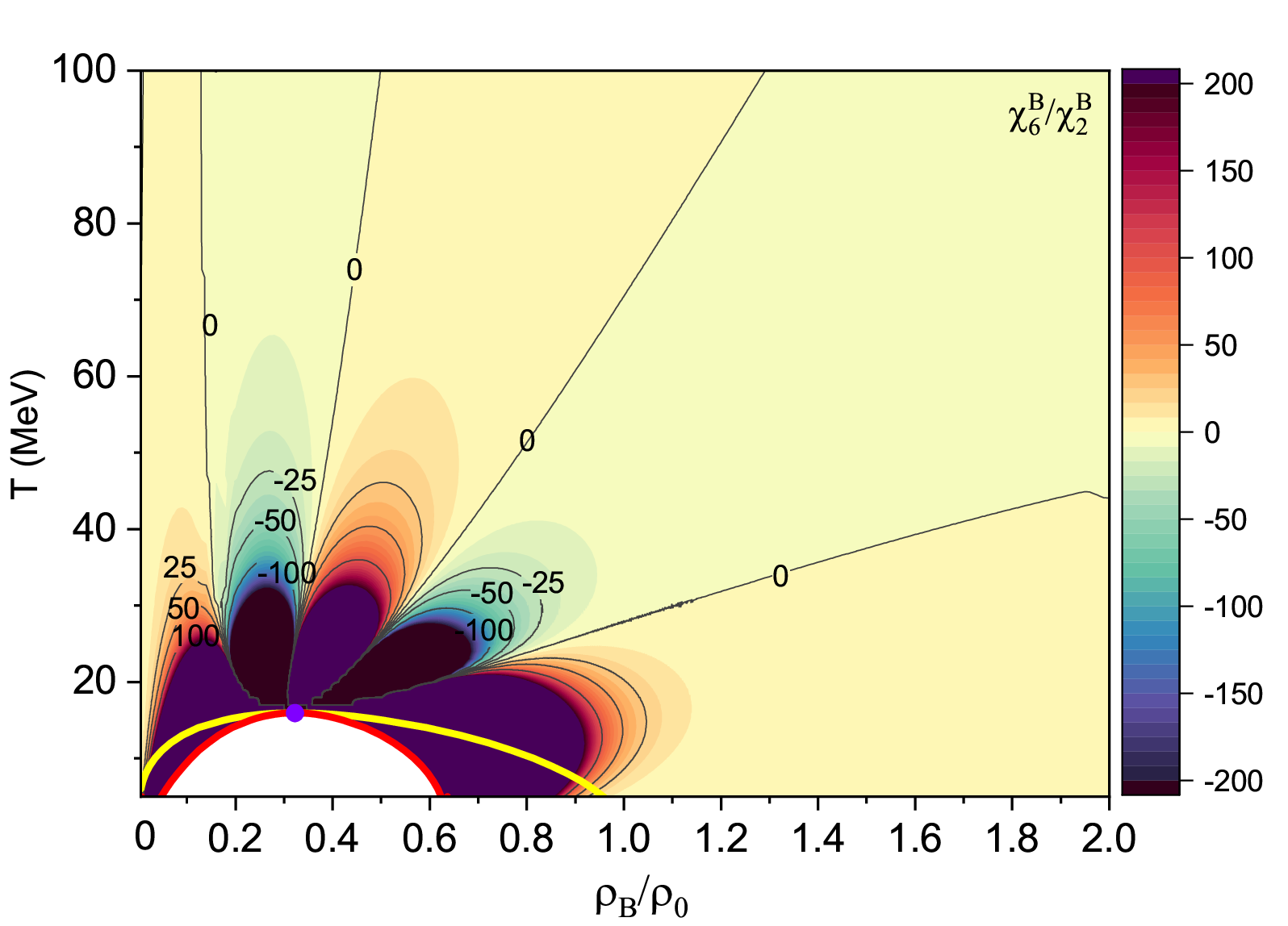}
\caption{(color online) Contour plots of the ratios of net baryon number susceptibilities, $\chi_3^B/\chi_1^B$, $\chi_4^B/\chi_2^B$, 
$\chi_5^B/\chi_1^B$ and $\chi_6^B/\chi_2^B$, in the $T-\rho_B$ phase diagram.
} 
\label{fig:6}
\end{figure*} 
In order to visualize the organization structure of density fluctuations in the phase diagram, we present the contour plots of net baryon number fluctuations in Figs.~\ref{fig:5} and \ref{fig:6}. The two figures explicitly demonstrate that the ratios of hyperorder susceptibilities exhibit more pronounced density fluctuations than those of lower-order ones. At the meanwhile, we can see that the closer the density fluctuations are to the  critical endpoint of the phase transition, the more frequently the oscillatory behavior of the density fluctuations appears. Additionally, in Fig.~\ref{fig:6}, we also present the net baryon number density fluctuations in the metastable phase. It can be observed that the ratios of net-baryon susceptibilities  change continuously from the stable phase to the metastable phase. On the boundary of first-order liquid-gas phase transition (excluding the critical point), the density fluctuations have finite values, while the density fluctuations tend to diverge on the spinodal line~(red line).

These behaviors of net baryon number fluctuations can be simply understood according to the relationship between $P/T^4$ and $\mu_B/T$. For each given temperature above the critical endpoint, $P/T^4$ increase monotonously as a function of $\mu_B/T$. For a lower temperature, the slop of $P/T^4$ increases faster with $\mu_B/T$ near the region where $\partial{\sigma}/\partial{\mu_B}$ changes rapidly, and the maximum slope is also larger. Such a feature results in that the higher order susceptibility oscillates much sharply in a relatively narrow range at lower temperature. The center of oscillation moves roughly along Line A (shown in Fig.~1) towards the CEP of liquid-gas phase transition.

To improve the theoretical calculation in the future, more degrees of freedom of hadrons and interactions should be included at relatively higher temperature. 
On the experimental side, we should keep in mind that the initial volume fluctuation effects in experiments become significant due to lower charged particle multiplicity at lower-energy collisions. 
Nevertheless, this study qualitatively predicts the behavior of net baryon number fluctuations at chemical freeze-out at low temperatures. 
With the increase of statistical data and the improvement of data precision in the second phase of beam energy scan at RHIC STAR, the deep combination of theory and experiment will bring important opportunities for exploring the phase structure of strongly interacting matter.

\section{summary and conclusion}

Fluctuations of conserved charges are sensitive probes to explore the phase transition of strongly interacting matter. In this research, using the non-linear Walecka model we  investigated the fluctuations of net baryon number up to sixth order caused by hadronic interactions in nuclear matter, and explored the relationship between hyperorder baryon number fluctuations and nuclear liquid-gas phase transition, including the stable and  metastable phases as well as the region far from the phase transition.

The calculation indicates that the fluctuations of net baryon number gradually increase from the high-temperature region to  critical region of nuclear LGPT. In particular, the fluctuations are prominent near the region where the $\sigma$ field or the nucleon mass changes rapidly for each temperature. It exhibits similar behavior to the chiral phase transition of quark matter. This can be further attributed to that the two phase transitions have the same universal class. 

Compared with the kurtosis and skewness of net baryon number fluctuations, the values  of hyperkurtosis~($\chi_6^B/\chi_2^B$) and hyperskewness~($\chi_5^B/\chi_1^B$) are more dramatic, which provide more sensitive signals to detect the phase transformation.
In combination with heavy-ion collision experiments, we further extracted density fluctuations at chemical freeze-out, and obtain the variation trend of baryon number fluctuations with the decrease of temperature (collision energy). With the release of more precise data in BES II program in the future, the results obtained in this research can be referred to analyze the QCD phase transition and nuclear liquid-gas phase transition. In addition, more physical mechanisms, such as the initial volume fluctuation effects, the non-equilibrium evolution and more hadronic degrees of freedom with complex interactions,  need to be considered to better understand the evolution of heavy-ion collision experiments at relatively lower collision energy. 

\section*{Acknowledgements}
This work is supported by the National Natural Science Foundation of China under
Grant No.~11875213.

\end{document}